\documentclass{sig-alternate}

\def\sym#1{\ifmmode^{#1}\else\(^{#1}\)\fi}
\usepackage{graphicx}
\usepackage{color}
\usepackage{amssymb}
\usepackage{epstopdf}
\usepackage{longtable}
\usepackage[lined,boxed,commentsnumbered]{algorithm2e}
\usepackage{tabularx}
\usepackage{balance} 
\usepackage{multirow}
\usepackage{amsmath}

\newcommand{\etal}{{et al.\ }}

\begin{document}

\title{Does a Daily Deal Promotion Signal a Distressed Business? An Empirical Investigation of Small Business Survival}
\numberofauthors{3} 
\author{
\author{ Ayman Farahat and Nesreen Ahmed and Utpal Dholakia }
\alignauthor Ayman Farahat\\
       \affaddr{Adobe Inc.}\\
       \email{farahat@adobe.com}
\alignauthor Nesreen K. Ahmed\\
       \affaddr{Purdue University}\\
       \email{nkahmed@purdue.edu}
\alignauthor Utpal Dholakia\\
       \affaddr{Rice University}\\
       \email{dholakia@rice.edu}
}

\date{}

\maketitle
\begin{abstract}
In the last four years, daily deals have emerged from nowhere to become a multi-billion dollar industry world-wide. Daily deal sites such as Groupon and Livingsocial offer products and services at deep discounts to consumers via email and social networks. As the industry matures, there are many questions regarding the impact of daily deals on the marketplace. Important questions in this regard concern the reasons why businesses decide to offer daily deals and their longer-term impact on businesses. In the present paper,  we investigate whether the unobserved factors that make marketers run daily deals are correlated with the unobserved factors that influence the business, In particular, we employ the framework of seemingly unrelated regression to model the correlation between the errors in predicting whether a business uses a daily deal and the errors in predicting the business' survival. Our analysis consists of the survival of 985 small businesses that offered daily deals between January and July 2011 in the city of Chicago. Our results indicate that there is a statistically significant correlation between the unobserved factors that  influence the business' decision to offer a daily deal and the unobserved factors that impact its survival. Furthermore, our results indicate that the correlation coefficient is significant in certain business categories (e.g. restaurants).
\end{abstract}
 
   
\section{Introduction}
\label{introduction}
 Daily deal sites such as Groupon represent a novel approach to Internet marketing
that tap into local markets, and based on the massive scale and rapid growth of
such sites, the business model has gained rapid adoption from a wide range of businesses. 
 
 Despite this success, a vocal contrarian view of the daily deals model has emerged. Its chief
criticism is probably skepticism about the value of daily deals to the merchants whose goods and
services are promoted. To be sure, enthusiastic advocates are easy to find; Groupon has claimed that
97\% \cite{Gupta}  of businesses using its service want to be featured again. But an independent study estimates
repeat intent at only 48.1\%. Some anecdotal reports are surprisingly harsh, including a highly
publicized blog posting by the owner of a New York bakery cafe, who described her Groupon
promotion as "the single worst decision I have ever made as a business owner".
 
The diverging views about the profitability and long term impact of the multi-billion  daily deals industry calls for a  thorough study and  more details than previous attempts. 
Any study pertaining to evaluate the impact of daily deals on key business metrics needs to isolate  the {\em causal impact } of  daily deals  from 
other factors that might be correlated with daily deal adoption and also impact business metrics.

In this paper, we examine the impact of daily deals  on the business  from the lens of modern econometrics. 
It is well known that the gold standard for evaluating a treatment effect such as daily deals is through randomized experiments. In practice however, it is not possible
to run the experiments needed to isolate the impact of daily deals from other confounding factors. Therefore, most studies including ours will have to contend with observational data. 
Working with observational  data poses its own set of challenges and it is exactly these challenges that call for tools from modern econometrics. 

One of the main  challenges of  working with observational data is unobserved heterogeneity. While two businesses might look identical along the observed dimensions such as category, location..{\em etc}, they may differ
along dimensions that can have a significant impact on the business. For example, a struggling business with a high staff turnover will be more likely to fail than a similar business. Also, struggling businesses might be more tempted to use a daily deal to help shore sales. A naive application of standard statistical techniques will show  daily deal adoption to be a factor in business failure while in reality it is staff turnover that contributed to the failure.
To alleviate the problems of unobserved heterogeneity, we treat the daily deal adoption as a {\em dependent variable}. A business's decision to adopt a daily deal will depend on a number of observed independent variables such as the business category, how popular are deals in category, etc. The daily deal decision will also be impacted by  unobserved factors. Similarly, business failure will  depend on observed dependent variables and unobserved factors. 
For example, consider a business that runs a daily deal despite of having no  apparent reason to do so. Does that signal anything about the unobserved reasons that might lead the business to fail down the line?
If the unobserved factors that make a business run a daily deal are correlated with the unobserved factors that lead to business failure then indeed the daily deal adoption conveys additional information about the business and should be taken into account. On the other hand, if there is no correlation between the unobserved factors in daily deal and failure, then knowing whether  a business offered a daily deal does not convey additional information. 

We resort to techniques from modern econometrics to help test  whether the unobserved factors in daily deal adoption  and failure are correlated. In particular, we use the framework of multiple equation models. One equation in our setting models  the business's decision to offer a  daily deal   and the second equation models the business failure. In order to model both equations, we need to identify the dependent variables (daily deal adoption  and failure) and a set of covariates that best describe each model. 

In this work, we tried to insure that our results were statistically  significant, robust to modeling assumptions and scalable. To help achieve statistical significance, we compiled a large data set that had information on daily deal adoption, business failure and information about the businesses. To test robustness of our models, we used two frameworks that make different  modeling assumptions. To  insure scalability, we developed a data driven approach that does not require expensive interviews with customers or lab experiments.  In particular, we crawled Yelp to get information about the business and whether a business was closed. For daily deal adoption, we used a data set from \cite{Byers2012}.

In this paper we develop a model for business failure in Chicago  where both the failure data and business information are derived from Yelp.  We also develop a separate model for daily deal adoption  where we join the data set from \cite{Byers2012} with Yelp data. We decided to use Chicago because in addition to being the third largest city in the U.S,   it is the home town of the largest daily deal provider ``Groupon" and had a large number of business adopting daily deals. We then developed joint models  of daily deal adoption and business failure. Our results show that a joint model of business failure and daily deal adoption  does a better job in explaining the data than the two separate models. We tested the robustness of our results by specifying two different modeling paradigms with different model assumption ; bivariate probit and seemingly unrelated regression. 

Our results show that models of business failure and  daily deals adoption using Yelp based features provide good performance.  Furthermore, we  find  a positive and statistically significant  correlation between the unobserved factors that make a business offer a daily deal and the unobserved factors that contribute to failure. Our results indicate that the correlation is strongest in case of Restaurants  {\bf 0.281}  and smaller in case of Spas {\bf 0.24}.   These results are consistent with results of Gupta \cite{Gupta} that found that  deals are generally good for Spas and  bad for Restaurants.
\vspace{1mm}

In summary, the main contributions of this paper are the following:
\vspace{-1mm}
\begin{itemize}
\item We conducted a data driven  large scale study to test and quantify whether the unobserved factors that make a business decide to make daily deals are correlated to the unobserved factors that impact the business survival.
\item We developed and analyzed  business survival and daily deal adoptions  models based on business features collected from Yelp.
\item We conducted statistical tests and two different econometrics frameworks to insure statistical significance of results.
\item Our results are consistent with findings from previous research work that used labor intensive surveys. 
\end{itemize}
This paper is organized as follows. In Section~\ref{sec:related}, we survey related work. In Section~\ref{dataset}, we describe our data and the methods employed to collect it. In Section~\ref{sec:econometric_model}, we  describe the necessary background from econometrics that is needed to develop our models. In Section~\ref{sec:yelpmodel}, we describe our experiments, results and evaluations. Finally, in Section~\ref{future} we summarize our findings and give directions for future work.
\section{Related Work}
\label{sec:related}

\begin{table*}
\begin{center}
\begin{tabular}{|| l | l | l | l||} 
\hline
\textbf{Business Category} & \textbf{Total No.} &  \textbf{Multiple Deals (\%)} & \textbf{Closed (\%)} \\
\hline
\hline
\textsc{Restaurants \& Bars} & 337 & 7.1 & 8.3 \\
\hline
\textsc{Beauty \& Spas} & 189 & 10 & 4.7 \\
\hline
\textsc{Active Life} & 151& 9.2 & 1.9\\
\hline
\textsc{Shopping} & 116  & 6.4 & 0.9\\
\hline
\textsc{Fitness \& Instruction} & 87 & 5.3 & 0.0\\
\hline
\textsc{Food} & 84 & 7.1 & 11.9\\
\hline
\textsc{Health \& Medical} & 77  & 12.1 & 0.0\\
\hline
\textsc{Nightlife} & 73 & 5.5 & 1.4\\
\hline
\textsc{Hair Salons} & 73 & 9 & 6.5\\
\hline
\textsc{Arts \& Entertainment} & 72 & 6.8 & 1.4\\
\hline
\end{tabular}
\vspace{-1mm}
\caption{Statistics of Groupon Businesses in Chicago (Jan-July 2011)} 
\vspace{-6mm}
\label{tab:groupontbl}
\end{center}
\end{table*}

Recently, there has been an increasing number of both empirical and theoretical research on daily deals or what is sometimes called voucher discounting. Most of this work focused on studying Groupon and LivingSocial discounted deals. For example, Dholakia studied the question of whether Groupon promotions are profitable for businesses and which businesses fare the best and worst after offering a Groupon promotion \cite{utpal_a,utpal_b}. He found mixed results where  some business owners reported their Groupon promotion was profitable and others regretted making the promotion based on their experience of lower spending and return rates from Groupon users. In another study, Arabshahi provided a detailed analysis that explains the Groupon business model and its underlying principles \cite{arabshahi}. In his paper, he explained that the main challenge facing merchants lies in identifying price-sensitive potential customers and offering them discounts. Therefore, Groupon can help merchants to apply price discrimination through the "highly discounted deals" provided to a massive scale of price-sensitive subscribers. In a similar work, Edelman {\em \etal} provided a theoretical study of the economics of Groupon deals from the perspective of participating merchants rather than from the perspective of the deal service provider \cite{edelman}. Their results indicate that voucher discounts are naturally good fits for certain types of merchants, and poor fits for others. 

Similarly, Gupta  {\em \etal} \cite{Gupta} investigated when are daily deals profitable for business by interviewing over 2000 business that offered daily deal through ``Groupon".  They  found that the success of a daily deal is far from certain and that the  return on investment varies widely. They identified the  types of businesses as a reliable predictor of profitability and that daily deals are good for spas  but bad for restaurants. 

The work presented in \cite{utpal_a,utpal_b} and \cite{Gupta} can be viewed as complementary to our work; while we focused  on a data driven scalable approach,  \cite{utpal_a,utpal_b} and \cite{Gupta} focused on a more labor intensive interview process. These two approaches are not exclusive with the findings from one guiding the other. 

Also, Ye {\em \etal} studied the group purchasing behavior of daily deals in Groupon and LivingSocial and they proposed a predictive dynamic model for group buying behavior \cite{ye}. Their model was able to predict the popularity of group deals as function of time. They also found that the different incentive mechanisms applied in Groupon and Living Social (individual threshold versus collective threshold) lead to different propagation behavior, which finally lead to different predictability. 

While studying daily deals is interesting in itself, another trend of research started to study the marriage between daily deal sites and the growing consumer phenomena such as Yelp. Byers {\em \etal} initiated the study of how daily deal sites affect the reputation of the business and in particular the business Yelp reviews \cite{byers_a}. In their first research paper, the authors studied the interplay between social networks and daily deal sites. They found that daily deal sites benefit from significant word-of-mouth effects during sales events. They also studied the effects of daily deals on the long-term reputation of merchants, based on their Yelp reviews before and after they run a daily deal. They found that the Yelp ratings of Groupon-bearing consumers were on average 10\% lower than those of their peers.

In another study, Byers {\em \etal} rigorously evaluated various hypotheses about underlying consumer and merchant behavior to understand the Groupon effect on businesses  \cite{byers_b}. They examined a number of hypotheses to justify the Groupon effect. For example, they illustrated a poor business behavior, and Groupon user experimentation to be possible root causes of the Groupon effect. They also found an evidence that on average Groupon users are no more critical than their peers. 

Similarly, Zervas tried to establish basic facts regarding the evolving quality of the deals that Groupon offers \cite{zervas}. He used Yelp ratings as a proxy for measuring the quality. Using simple regression analyses, Zervas found a statistically significant negative correlation between the time deals that have been offered and the Yelp ratings of the merchants who offered them. Further, he discussed some possibilities that might cause these trends. For example, as Groupon is expanding the number of deals it offers, it has to work with some lower-rated merchants. Also, it is possible that better-rated merchants dropping out of running Groupon deals, and Groupon has to substitute them for merchants with some lower-rated merchants.  

Our work builds on \cite{byers_b} and \cite{zervas} by explicitly modeling the decision to offer a daily deal and leveraging the error in the model to explain part of the unobserved factors in modeling business performance. 

Business reviews collected from Yelp have  also been studied by Luca \cite{Luca2011}. Luca evaluated the impact of Yelp reviews on restaurant's quarterly earnings in Seattle using the framework of regression discontinuity. 
Luca finds that the observed response to Yelp rating are consistent with bayesian learning.   Under the bayesian hypothesis, reactions to signals are stronger when the signal is more precise (i.e., the Yelp average rating contains more information when the number of reviews is high). Moe precisely, a change in a restaurant's average rating has $50\%$ more impact when the restaurant has at least $50$ reviews (compared to a restaurant with fewer than $10$ reviews). 

Luca \cite{Luca2011} also tests whether restaurants are gaming the rating system using the 
McCrary \cite{McCrary2008} test. The intuition of the test is as follows. Suppose that restaurants were gaming
Yelp in a way that would bias the results. Then, one would expect to see a disproportionately
large number of restaurants just above the rounding thresholds. Luca \cite{Luca2011} finds that this is not the case. 
The results presented in \cite{Luca2011} are related to our work in two ways. First, we use the concept of ``Bayesian Learning" to derive statistically significant predictors of both daily deal adoption and business failure. Second, the McCrary test \cite{McCrary2008} suggests that the reviews on Yelp truly reflect the opinion of the Yelp community and are not being manipulated by the businesses on Yelp.  

Pindyck and Rubinfeld \cite{PindyckRubinfeld} model the relation between private school attendance and voting for property tax increases that are used in part to finance public schools. In this application, the variables are whether children attend private school, number of years the family has been at the present
residence , log of property tax , log of income and whether the head of
the household voted for an increase in property taxes. Pindyck and Rubinfeld \cite{PindyckRubinfeld} wanted to test the hypothesis that parents of children who attended private school will have no incentive from an increase in property taxes that finance public schools and will vote against any such increases.
 Pindyck and Rubinfeld  \cite{PindyckRubinfeld} model the bivariate  outcomes of whether children attend private school and whether
the head of the household voted for other covariates. 
They conclude that the two outcomes are independent and that the  voting patterns of parents of children attending private schools do not differ from parents of children attending public schools. 
Our work is related to  \cite{PindyckRubinfeld} in that we test for the independence of two binary outcomes; daily deal adoption and business failure.

\section{Data Collection \& Analysis}
\label{dataset}

\begin{table}[h]
\small
\begin{tabular}{|| l| l| l ||} 
\hline
\textbf{Business Category} & \textbf{Total No.} &  \textbf{Closed (\%)}  \\
\hline
\hline
\textsc{Restaurants \& Bars} & 8490 & 17.7  \\
\hline
\textsc{Shopping} & 4961  & 10.6 \\
\hline
\textsc{Food} & 3259 & 14.9  \\
\hline
\textsc{Beauty \& Spas} & 2692 & 5.9  \\
\hline
\textsc{Health \& Medical} & 2666  & 2.14 \\
\hline
\textsc{Nightlife} & 1851 &  16.0 \\
\hline
\textsc{Active Life} & 1301 & 6.3 \\
\hline
\textsc{Arts \& Entertainment} & 1267 & 6.0 \\
\hline
\textsc{Hair Salons} & 948 & 5.3\\
\hline
\textsc{Fitness \& Instruction} & 740 & 7.6 \\
\hline
\end{tabular}
\vspace{-1mm}
\caption{Statistics of Yelp Businesses in Chicago as of July 2012}
\label{tab:yelptbl}
\vspace{-3mm}
\end{table}

Our dataset collection has two major components. First, we collected data from Groupon as one of the top deal sites that offers daily deals in Chicago. Second, we collected data from Yelp for all the businesses in Chicago.  

\paragraph{Groupon Data}We used the Groupon data set compiled by Byers \etal \cite{byers_a} which includes the web links of $16,692$ deals offered by Groupon in $20$ U.S. cities between January and July $2011$. In this paper we focus only on the subset of Groupon deals offered in Chicago. We selected Chicago not just because it is the third largest city in U.S but also it the home town of Groupon. When the Groupon business is featured on {\em Yelp}, Groupon occasionally uses that information to promote the deal by including a link to the Yelp site as well as other information (e.g. star rating and selected customer reviews). However, in some cases Groupon does not mention the Yelp link on the deal page even if the business has a Yelp link.

We are interested in Yelp since it provides a wide range of information about the business. For Example, Yelp provides business location, number of reviews, date of review, star ratings, review text and other features such as alcohol license and price range. Moreover, Yelp indicates whether the business is still in operation or whether the business has closed by adding the string "CLOSED" next to the business name. Previous research has shown the potential of Yelp to indicate business parameters and performance \cite{Luca2011} as we explained in section~\ref{sec:related}.

Groupon provides a convenient API \footnote{http://www.groupon.com/pages/api} to collect information about the deals, however, we decided to develop our web crawler to extract features that are not supported by the API. For example, whether there is a link to Yelp or not. We initially had $1861$ Groupon deals from chicago, with approximately $60\%$ of them had their Yelp links listed. For the deals without Yelp links, we used the Yelp search feature to find a match for the Groupon business on Yelp. Specifically, we searched Yelp by the business name and the zip code listed on the deal webpage. Typically, Yelp return search results for relevant matches within the given zip code and other nearby zip codes. However, we report only the query results that exactly matched both the business name and its zip code. By the end of this matching process, we successfully associated $1184$ Groupon deals with Yelp links, we call them {\em "GrouponDealsWithYelp"}. We also observed that some businesses offered multiple deals while others offered only one deal. While these deals are supposed to be all in Chicago, we found few cases where the deal zip code was outside Chicago (for other branches of the business in other states). Since we focus only on Chicago business we decided to filter these cases. Finally, we developed our web crawler to extract the Yelp information for the businesses in the set {\em "GrouponDealsWithYelp"}. After filtering the businesses that had zero reviews (since they don't provide any information about the business), we observed $985$ businesses with Groupon deals and Yelp links. We call this set {\em GrouponBusiness"}. Table~\ref{tab:groupontbl} provides statistics of the set {\em "GrouponBusiness"} for the top $10$ business categories.

\paragraph{Yelp Data}We crawled the Yelp site to collect all the businesses that appear in Chicago (regardless whether they offer a deal or not). Yelp uses a structured format that arranges business names by alphabetical order. We initially had $38,000$ businesses listed in Yelp. After filtering all the cases that have zero reviews, we had $32,424$ with approximately $9\%$ failed businesses (closed) and approximately $4\%$ offered Groupon between January and July 2011. We refer to this set as our {\em Yelp Population} to represent the real population of businesses in Chicago. However, we expect Yelp data would represent certain business categories (e.g. restaurants) more than others (e.g. Insurance). Therefore, we decided to analyze only the top business categories. In Table~\ref{tab:yelptbl} we provide statistics of the dataset we collected from Yelp for the top $10$ business categories. 

In the next sections, We proceeded to build a model for predicting failure using Yelp data. The details of the model are shown in Section \ref{subsec:survival_probit}. To build the bivariate model of Groupon adoption and business failure, we had to restrict our analysis to the businesses that did not offer a Groupon but were operating during the same period as the {\em "GrouponBusiness"}. We use the date of the last review posted for the business as a proxy for the closing date. We refer to the set that includes all of the businesses that did not offer Groupon deals and did not fail before January 2011 as {\em "nonGrouponBusiness"}. 

\paragraph{Analysis}

Table~\ref{tab:groupontbl} provides statistics of the set {\em "GrouponBusiness"} for the top $10$ business categories. Also, Table~\ref{tab:yelptbl} provides statistics of our data {\em Yelp Population} which includes both the two sets {\em "GrouponBusiness"} and {\em "nonGrouponBusiness"}. From Table~\ref{tab:groupontbl} and Table~\ref{tab:yelptbl}, we observe some interesting patterns. First, we observe a difference in the ranks of the business categories between the total population of Yelp data and the {\em "GrouponBusiness"} except for the first category "Restaurants \& Bars". We conjecture that some business categories could be popular on Yelp but they don't have the incentives to offer daily deals. We also observe that the highest percentage of closed businesses comes from the categories "Restaurants \& Bars", "Food", and "Nightlife".
In addition, we observed that some business in the set {\em "GrouponBusiness"} had the incentives to offer multiple daily deals during the six month period we analyzed.

\begin{figure}[htbp]
\centering
\includegraphics[width=0.45\textwidth]{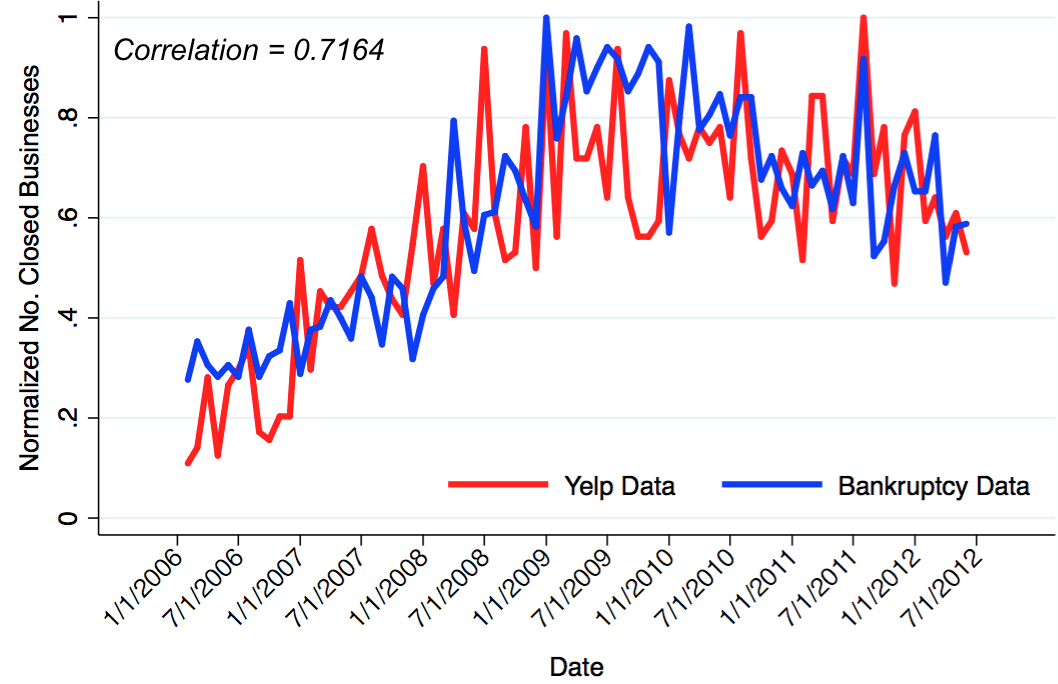}
\vspace{-2mm}
\caption{Normalized Yelp Closed Businesses versus Normalized Bankrupted Businesses between Jan 2006 and July 2012}
\label{fig:cl2}
\end{figure}

As we discussed in section~\ref{sec:related}, previous research work emphasized the potential of Yelp as a proxy measure for business key performance indicators (e.g. survival, consumer appeal, and revenue) \cite{Luca2011,byers_a,byers_b,zervas}. Luca in \cite{Luca2011} found that $69\%$ of restaurants in Seattle are listed on Yelp. Also, Luca showed that changes in Yelp ratings are associated with changes in revenues \cite{Luca2011}. These studies indicate the potential of Yelp data as representative of the true population. However, to the best of our knowledge, none of this research analyzed Yelp as a source of business failure information. 
Therefore, to test the representativeness of Yelp data as a source of business failure information, we compared the number of closed businesses collected from Yelp to the number of bankrupted businesses as reported by bankruptcy filings of the Northern Illinois (which includes Chicago)  open court records collected by the bankruptcy data project at Harvard \footnote{http://bdp.law.harvard.edu}. Both the data from Yelp and court bankruptcy filings are  between January 2006 and July 2012. Figure~\ref{fig:cl2} shows the plot of the two normalized time series data . As shown in the plot, there is a strong correlation between the number of closed businesses computed from Yelp and the number of bankrupted businesses (correlation coeff. $=0.7164$). Also, the two datasets have a similar trend as shown in Figure~\ref{fig:cl2}.       

\vspace{-2mm}

\section{Econometric Framework}
\label{sec:econometric_model}

In this section, we develop the econometric model. To motivate the need for an econometric model, consider the task of deciding whether running a daily deal increases the risk of business failure. We can model this as a regression problem of the form 
\begin{equation}
failure_i  = \beta * dailydeal_i + e_i  \label{eqn:naive}
\end{equation}
where "$failure_i$" is whether business "i" fails,  , "$dailydeal_i"$  indicates whether the business ran a daily deal  and $e_i$ is the error term and captures all the factors that are not included in the model. These factors are referred to as the "unobserved heterogeneity". The unobserved heterogeneity can include factors such as  location specific risk,  category risk .. .{\em etc}. It can also include factors that are correlated with the business decision  to run a daily deal  such as whether the business is struggling. Consider for example the class of  struggling businesses that use daily deals  as a last resort to attract more customers. In expectation, a struggling business will have a higher risk of failure. However, if we were to perform a regression analysis using Equation \ref{eqn:naive}, we will over estimate the impact of daily deals ($\beta$) because a large number of the struggling businesses that used daily deals failed. While in fact, these businesses did not fail because of daily deals, they failed because of internal problems that happened to be correlated with the decision to offer a daily deal. In this case, we have an {\em omitted variable bias} \cite{Green6}. The omitted variable bias will only arise however, if the omitted variable (which in our case is whether the business was struggling), is correlated with one of the regressors ( the decision to use daily deal). 

The problem can also be viewed from the lens of endogeneity. A perquisite to any regression model is that the regressors are exogenous,  \cite{Green6}  {\em i.e.,} the regressor variable comes from outside of the model and cannot be explained by any of the variables of the model. However, as we have seen in the above example, the regressor $dailydeal$ was correlated with the errors.  Consequently, the variable $dailydeal$ is not exogenous as it it can be explained in part by the errors. The omitted variable bias is a form of endogeneity that results in a biased estimator. 

If we wanted to estimate the impact of daily deals on the business survival, we will need to find an {\em instrument} \cite{Green6}. An instrument is a variable that is correlated with the endogenous variable (dailydeal) but is not correlated with error (struggling business). For example, we could use the size of the daily deals providers sales force as an instrument since it is likely to be correlated with decision to adopt daily deals but unlikely to be correlated with the error term. A model with an {\em instrumental variable} is estimated using {\em 2 stage least squares (2sls)}. In future work, we will address the use of instrumental variables approach. 

However, in this paper, we don't directly attempt to model the impact of the daily  deal on business survival. Instead, we focus on assessing the correlation between the unobserved factors that make a business offer a daily deal and the unobserved factors that influence failure.  In that vein, we  consider two different econometric models. In section \ref{subsec:bivariate_model}, we use a bivariate probit model to check whether the errors  in predicting if the business decision to run a daily deal are  correlated with the  errors in predicting the business failure. 
Since the errors model the unobserved heterogeneity, the bivariate model will help identify whether the unobserved
factors that make a business run a daily deal are also correlated with the unobserved factors that contribute to the
business failure. In section \ref{subsec:sure_model}, we relax the conditions that
the dependent variables (run a daily deal and failure) are binary and use the framework of seemingly unrelated regression " (SURE) to answer the same question addressed by the bivariate profit. 

Our decision to use  the SURE model is motivated by two factors. First, the bivariate probit model makes the explicit assumption that  the error terms are jointly normal. While this might sound like a reasonable assumption, we have no way of checking its validity. On the other other hand, the SURE model does not make any assumptions about the joint distributions of the error, but rather uses the variance of these distributions to derive its estimate. Therefore, the use of two models provides us with a  robustness test as in \cite{TuckerGoldfarb2011}. The second reason we consider the SURE model is that  recent developments in econometrics \cite{Harmless} that used actual experimental data  do not show any advantages of enforcing the limited range of dependent variables.

An important question is whether a single model best  explains the two dependent outcomes (daily deals adoption and  survival ) or whether we need two separate models. In section \ref{subsec:spec_testing}, we describe a specification test based on the likelihood of the data and show how they relate to parametric technique, that  test the  null hypothesis of no correlation between the error terms.
 
\subsection { Bivariate Probit}
\label{subsec:bivariate_model}
Assume that we have a random sample of $N$ observations where each observation is donated by $i$ such that $i={1,...,N}$.    
In ordinary regression models, we typically observe only one dependent variable for each observation $Y=(Y_{1},...,Y_{N})$. However, in  the general case general case, we can observe multiple dependent variables for each observation. 
Let $Y_{ji} $ denote the response of the $i^{th}$ observational unit for the $j^{th}$ dependent variable. A typical situation is the case when we observe $2$ variables, such that $Y_{1i} \text{ and } Y_{2i}$ are two  binary dependent variables. 

Traditional probit models can generally be described as latent variable models, in which we define a latent variable $Y^*$  such that $Y =1_{(Y^*>0)}$. In this section, we consider the bivariate probit model.
The bivariate probit model belongs to the generalized class that is usually used to estimate several correlated binary variables jointly. These often represent two interrelated decisions, for example, to adopt two different, but related, policy initiatives. In the bivariate probit model, we have two separate probit models with correlated error terms. Specifically, we have two binary dependent variables for each $i^{th}$ observational unit : $Y_{ji}={Y_{1i},Y_{2i}}$ such that $j=1,2$. Therefore, we have the following model
\begin{align}
Y_{1i}^*&= X_{1i}\beta_1+\varepsilon_{1i}\\
Y_{2i}^*&=X_{2i}\beta_2+\varepsilon_{2i}
\end{align}
where $Y_{1i}^*$ and $Y_{2i}^*$ are the latent variables and they are related to the dependent variables by the following equation
\begin{align}
Y_{1i} &=1_{(Y_{1i}^*>0)}\\
Y_{2i}&= 1_{(Y_{2i}^*>0)}
\end{align}
The vectors $ X_{1i}$  denotes the $[N_1  \times 1]$ vector  of exogenous regressor for dependent variable "1" . $ X_{2i}$  denotes the $[N_2  \times 1]$ vector  of exogenous regressor for dependent variable "2" .
In the bivariate probit model, the main assumption is that error terms $\varepsilon_{1i}, \varepsilon_{2i}$ are independent across observations ``$i$" but may have cross-equation correlations. Therefore, we have $E[\varepsilon_{ji}\varepsilon_{jk}|x ]= 0  
\text{ } \forall i \neq k $. In addition, the error terms are drawn from a bivariate normal distribution \cite{Green6}
\begin{equation}
\left(
\begin{array}{c}
\varepsilon_{1i}\\
\varepsilon_{2i}
\end{array}
\right)|X
\sim \mathcal{N}
\left(\left(
\begin{array}{c}
0\\
0
\end{array}
\right)
,
\left(
\begin{array}{cc}
1&\rho\\
\rho&1
\end{array}
\right)\right)
\end{equation}

Where $\rho$ is a correlation parameter denoting the extent to which the two $\varepsilon$'s covary. The conditional expectation for the bivariate normal distribution is given by 
\begin{equation}
E(\varepsilon_{2i} | \varepsilon_{1i}> z) = \rho  {\phi_1(z) \over \Phi_1(-z)}  \label{eqn:Mills}
\end{equation}
where  $ \Phi_1, \phi_1$ are  the univariate normal cumulative distribution  and density functions respectively.Equation \ref{eqn:Mills} is the Inverse Mills ratio and has been used extensively in econometrics , \cite{Heckman79} \cite{Green6}. 
Equation \ref{eqn:Mills} shows that the in case of bivariate probit,  the errors are not independent, for example the error in estimating the probability of offering a Groupon deal corresponds in expectation to a large error in estimating the probability of business failures. The errors typically correspond to unobserved variables and by implication the Inverse Mills ratio indicates that these two variables move in synchronization. 
Fitting the bivariate probit model involves estimating the values of the parameters $\beta_1,\ \beta_2, and \ \rho $.
We use maximum likelihood estimation to estimate the parameters. The likelihood function $L$ is defined as:
\begin{eqnarray}
 \mathcal{L} & = &  \prod _{i \in N} P(Y_{1i}=1,Y_{2i}=1)^{Y_{1i}Y_{2i}} \nonumber \\
 & & P(Y_{1i}=0,Y_{2i}=1)^{(1-Y_{1i})Y_{2i}} \nonumber \\
& & P(Y_{1i}=1,Y_{2i}=0)^{Y_{1i}(1-Y_{2i})} \nonumber \\
& & P(Y_{1i}=0,Y_{2i}=0)^{(1-Y_{1i)(1-Y_{2i})}}
\end{eqnarray}
Substituting the latent variables  $Y_1^* $ and $Y_2^* $ in the Probability functions and taking the logarithm gives the log likelihood function $LL$:
\begin{eqnarray}
\mathcal{LL } & = & \sum_{i \in N}  Y_{1i}Y_{2i} \ln P(\varepsilon_{1i}>-X_{1i}\beta_1,\varepsilon_{2i}>-X_{2i}\beta_2) \nonumber \\
& & +(1-Y_{1i})Y_{2i}\ln P(\varepsilon_{1i}<-X_{1i}\beta_1,\varepsilon_{2i}>-X_{2i}\beta_2)   \nonumber \\ 
& & +Y_{1i}(1-Y_{2i})\ln P(\varepsilon_{1i}>-X_{1i}\beta_1,\varepsilon_{2i}<-X_{2i}\beta_2) \nonumber \\
& & +(1-Y_{1i})(1-Y_{2i})\ln P(\varepsilon_{1i}<-X_{1i}\beta_1,\varepsilon_{2i}<-X_{2i}\beta_2) \nonumber \\
\end{eqnarray}
After rearranging the terms, the log-likelihood function becomes:
\begin{eqnarray}
\mathcal{LL }  &=& = \sum_{i \in N}  Y_{1i} Y_{2i}\ln \Phi(X_{1i}\beta_1,X_{2i} \beta_2,\rho)  \nonumber \\
& & +(1-Y_{1i})Y_2\ln \Phi(-X_{1i}\beta_1,X_{2i}\beta_2,-\rho) \nonumber \\
& & +Y_{1i}(1-Y_{2i})\ln \Phi(X_{1i}\beta_1,-X_{2i}\beta_2,-\rho)  \nonumber \\
& & +(1-Y_{1i})(1-Y_{2i})\ln \Phi(-X_{1i}\beta_1,-X_{2i}\beta_2,\rho)  \nonumber \\
\label{eqn:likejoint}
\end{eqnarray}
Note that $ \Phi $ is the cumulative distribution function of the multivariate normal distribution|bivariate normal distribution and $\phi$ is the corresponding density function. $ Y_{1i} $ and $ Y_{2i} $ in the log-likelihood function are observed variables being equal to one or zero.

\subsection{Seemingly Unrelated Regression}
\label{subsec:sure_model}
In the previous section, we enforced the constraint  that the dependent variable have a limited range {\em limited dependent variables}. The bivariate probit frameworks  explicitly uses bivariate normal distribution to model the joint distribution of the error terms. To test whether this assumption is indeed valid, we relax these constraints and use  the framework of seemingly unrelated regression (SURE) proposed by Zellner \cite{Zellner62}.  Furthermore, there  is mounting evidence in the econometrics literature  \cite{Harmless} that argues in favor of using Ordinary least squares (OLS) even when the dependent variables are binary. 
\begin{align}
Y_{1i}&= X_{1i}\beta_1+\varepsilon_{1i}\\
Y_{2i}&=X_{2i}\beta_2+\varepsilon_{2i}
\end{align}
As before, the assumption of the model is that error terms $\varepsilon_{1i}, \varepsilon_{2i}$ are independent across observations "i"  but may have cross-equation correlations. Therefore we have $E[\varepsilon_{ji}\varepsilon_{jk}|x ]= 0  \text{ } \forall i \neq k $ and
\begin{equation}
\left(
\begin{array}{c}
\varepsilon_{1i}\\
\varepsilon_{2i}
\end{array}
\right)|X
\sim \mathcal{N}
\left(\left(
\begin{array}{c}
0\\
0
\end{array}
\right)
,
\left(
\begin{array}{cc}
{\sigma_{11}}^2&\rho{\sigma_{11}} {\sigma_{22}}\\
\rho {\sigma_{11}} {\sigma_{22}}&{\sigma_{22}}^2
\end{array}
\right)\right)
\end{equation}

\begin{equation}
\begin{pmatrix}y_{11} \\ y_{12} \\ \vdots \\ y_{1N} \\ y_{21} \\ y_{22} \\ \vdots \\ y_{2N} \end{pmatrix} = 
    \begin{pmatrix} X_{11} & 0 \\ X_ {12}  & 0 \\ \vdots&\vdots \\  X_{1N} & 0 \\  0&  X_{21}  \\  0 & X_ {22}   \\ \vdots&\vdots \\  0 &  X_{2N}    \end{pmatrix}
    \begin{pmatrix}\beta_1 \\ \beta_2  \end{pmatrix} +
    \begin{pmatrix}\varepsilon_{11} \\ \varepsilon_{12} \\ \vdots \\  \varepsilon_{1N} \\\varepsilon_{21} \\ \varepsilon_{22} \\ \vdots \\  \varepsilon_{2N}  \end{pmatrix}
 \end{equation}
Or in a more compact notation 
$ {\bf Y} = {\bf X} \beta + \varepsilon,  \qquad \mathrm{E}[\varepsilon|X]=0,\ \operatorname{Var}[\varepsilon|X]=\Omega.$ 

The SURE regression differs from the OLS in that the  covariance matrix is not spherical {\em i.e., }  $\operatorname{Var}[\varepsilon|X] \neq \sigma^2 I_n$ where $I_N$ is the identity matrix. 
In ordinary least squares, the Best Linear unbiased estimator  for the parameters $\beta $ is given by $\widehat \beta_{OLS} = (X' X)^{-1} X' y$. Once we have a non-spherical covariance matrix OLS is not efficient. 
To overcome this restriction, we use {\em feasible generalized least squares}  which is  a two stage estimator.  In the first stage, we run ordinary least squares estimation assuming that the two equations are independent. The residuals from the OLS are used to estimate the elements of the covariance matrix  $\hat\sigma_{ij} = \frac1 N\, \hat {\bf \varepsilon_i}^T \hat{ \bf \varepsilon_j} .$ 
In the second stage, we run {\em weighted  least squares} using the previously estimated covariance matrix. Feasible generalized least squares method estimates $\beta$  by minimizing the squared Mahalanobis length of the residual vector $ \hat\beta_{FGLS} = \underset{b}{\rm arg\,min}\,(Y-Xb)'\,\Omega^{-1}(Y-Xb)$ (note the in case of ordinary least squares $\Omega$ is diagonal  and therefore $ \hat\beta_{OLS} = \underset{b}{\rm arg\,min}\,(Y-Xb)'(Y-Xb)$ ). The explicit form of the estimator is given by 
\begin{equation}
\hat\beta_{FGLS} = (X'\Omega^{-1}X)^{-1} X'\Omega^{-1}Y.
\end{equation}
To test whether the two equations are best modeled using a SURE or can be modeled using ordinary least squares, it suffices to test whether the errors $\varepsilon_{1t}, \varepsilon_{2t}$ are correlated. The Breusch-Pagan test \cite{BreuschPagan} which is widely used in detecting  heteroskedasticity can be applied. 

\begin{table*}[ht]
\begin{center}
\begin{tabularx}{\textwidth}{||l l X||} 
\hline
\textbf{Symbol} & \textbf{Variable} & \textbf{Description} \\
\hline
\hline
$isClosed$ & \textsc{isClosed} & A binary outcome indicating whether the business failed (isClosed =1) or operating (isClosed = 0)  \\
\hline
$isGroupon$ &  \textsc{isGroupon} & A binary outcome indicating whether the business made a deal (isGroupon =1) or no deals (isGroupon = 0)  \\
\hline
$fzrisk$ & \textsc{Fail Ziprisk} & The percentage of failed businesses in the same zip code  \\
\hline
$fprisk$&\textsc{Fail Pricerisk} & The percentage of failed businesses in the same price category \\
\hline
$gzrisk$&\textsc{Groupon Ziprisk} & The percentage of businesses that made Groupon deals in the same zip code  \\
\hline
$gprisk$&\textsc{Groupon Pricerisk} & The percentage of businesses that made Groupon deals in the same price category \\
\hline
$rate$&\textsc{Rating} & Average Yelp Rating  \\
\hline
$nreview$&\textsc{Reviews} & Number of Yelp Reviews  \\
\hline
$price$&\textsc{Price Category} & \{1,2,3,4\} From cheap to most expensive \\
\hline
\end{tabularx}
\vspace{-3mm}
\caption{Description of Variables}
\vspace{-5mm}
\label{tab:vardesc}
\end{center}
\end{table*}
\vspace{-3mm}

\subsection{Specification Testing}
\label{subsec:spec_testing}
We use the Akaike  Information Criteria (AIC) \cite{Green6}  to test whether the data is best described by two separate models or a single joint model. In the case of the bivariate  probit model, the joint model will have one additional parameter (the correlation coefficient).  We therefore compare the log likelihood of the joint models  in Equation \ref{eqn:likejoint} to the log likelihood of the separate equations and test whether the difference is greater than ``1.0". 
In the case of   SURE model, we replace the log likelihood with the  the sum of squared residuals \cite{Green6}.
For the parametric approach, we employ  the  Cramer-Rao bound to compute parameter's mean and variance from the maximum likelihood estimator \cite{Green6}.  We use a ``t-test" to test whether the  parameters including the correlation coefficient $\rho$ are different from ``0". In the SURE setting, we use the Breusch -Pagan framework to test the correlation coefficient. 
\section{Experiments and Results}
\label{sec:yelpmodel}
In this section, we present the experimental setup and results of several regression models both in a univariate framework and a multiple equation framework. In the univariate framework, we use {\em a probit model} to analyze the significance of the different factors that may impact the business survival. Similarly,  we use {\em a probit model} to analyze the significance of the different factors that influence the business decision to make a daily deal in Groupon. In the multiple equations framework, we use bivariate probit and  seemingly unrelated regression frameworks to jointly model the business survival and the business daily deal decision. Here, our goal is to specifically investigate whether the unobserved factors that make the business offer a daily deal are correlated with the unobserved factors that impact the business survival. We conduct several statistical tests to test if the data is best described by a joint model and whether the  correlation between the unobserved factors  is significant. Our results indicate that joint models fit the data better than single univariate models and there is strong significant correlation for the two business categories restaurants and spas.

\subsection{Sampling}
When modeling  business survival and Groupon decision we need to take into account the relative frequency of both events. On one hand, the probability of a business failure in any given year is fairly low (in order of 8\%). On the other hand, daily deals is a relatively new phenomena which is still being evaluated, the number of businesses that leverage a daily deal are fairly low compared to the businesses that don't offer daily deals. 
In that vein, we are attempting to model two rare events : daily deal decision and business failure. Statistical models tend to underestimate the probability of rare events\cite{Green6}.  
Since the vast majority of the businesses will be non-groupon and non-closed, the model will assign a large negative constant to the two equations that model the daily deal and failure models. The large constants will make the error terms small and can impact the model's ability to detect the correlation between the error terms in the  two equations. 
To address this problem, we first apply the Bivariate Probit and the SURE to the entire data set. We then restrict the models such that there is no constant terms. We find that in the first case (unrestricted full sample), a positive but not statistically significant correlation between the error terms. In case of the full sample but  no-constant , we find that there is a strong positive correlation between the error terms. This confirms our intuition about the inability of the full data model to capture the two rare events. 
Therefore, we randomly sampled from the population of "{\em nonGrouponBusiness}" to account for the sparsity problems and selection biases that can be caused by data collection. 
\subsection{Univariate Probit Models}
A {\em probit model} is a type of regression used to model a binary dependent variable. Here, we define two binary dependent variables. First, we define the variable {\em isClosed} to represent whether the business has failed ({\em isClosed=$1$}) or the business is still operating ({\em isClosed=$0$}). Second, we define the variable {\em isGroupon} to represent whether the business has made at least one daily deal ({\em isGroupon=$1$}) at Groupon or the business did not make any deals ({\em isGroupon=$0$}). We develop a business survival model and Groupon decision model to analyze the factors that influence {\em isClosed} and {\em isGroupon} respectively.

\def\onepc{$^{\ast\ast}$} \def\fivepc{$^{\ast}$}
\def\tenpc{$^{\dag}$}
\def\legend{\multicolumn{4}{l}{\footnotesize{Significance levels
:\hspace{1em} $\dag$ : 10\% \hspace{1em}
$\ast$ : 5\% \hspace{1em} $\ast\ast$ : 1\% \normalsize}}}
\vspace{-2mm}
\begin{table*}[ht]
\parbox[c]{.5\textwidth}{
\begin{center}
\caption{Yelp Population Data}
\vspace{2mm}
\label{tab:tabres_closed_all}
\begin{tabular}{l r @{} l c }\hline\hline
\multicolumn{3}{l}
{Dependent Variable: isClosed} & {AUC = 0.674}\\\hline\hline
\multicolumn{1}{c}
{\textbf{Variable}}
& \multicolumn{2}{c}{\textbf{Coefficient}}  & \textbf{(Std. Err.)} \\ \hline
Fail Ziprisk  &  6.786&\onepc  & (0.317)\\
Fail Pricerisk  &  7.039&\onepc  & (0.380)\\
Rating $\times$ Reviews  &  -0.003&\onepc  & (0.000)\\
Reviews  &  0.008&\onepc  & (0.002)\\
Rating  &  -0.061&\onepc  & (0.010)\\
Intercept  &  -2.287&\onepc  & (0.060)\\
\hline
\legend
\end{tabular}
\end{center}
}
\parbox[c]{.5\textwidth}{
\begin{center}
\caption{Yelp Population Data}
\vspace{2mm}
\label{tab:tabres_groupon_all}
\begin{tabular}{l r @{} l c }\hline\hline
\multicolumn{3}{l}
{Dependent Variable: isGroupon} & {AUC = 0.894}\\\hline\hline
\multicolumn{1}{c}
{\textbf{Variable}}
& \multicolumn{2}{c}{\textbf{Coefficient}}  & \textbf{(Std. Err.)} \\ \hline
Groupon Ziprisk  &  4.521&\onepc  & (0.136)\\
Groupon Pricerisk  &  10.771&\onepc  & (0.888)\\
Rating $\times$ Reviews  &  -0.001&\onepc  & (0.000)\\
Reviews  &  0.006&\onepc  & (0.001)\\
Rating  &  0.057&\onepc  & (0.021)\\
Intercept  &  -2.848&\onepc  & (0.093)\\
\hline
\legend
\end{tabular}
\end{center}
}
\vspace{-4mm}
\end{table*}
\vspace{1mm}

\paragraph{Business Survival Model}
\label{subsec:survival_probit}
\vspace{1mm}
\noindent
We model the business survival variable {\em isClosed} as a probit function of a number of business variables we collected from Yelp (refer to Table~\ref{tab:vardesc} for notations):
\begin{equation}
isClosed  = f(rate \times nreview,  rate, nreview, fzrisk, fprisk) 
\label{eqn:survival_probit}
\end{equation}
We tried a number of other specifications and selected the specification  of Equation \ref{eqn:survival_probit} based on the AIC.
Table~\ref{tab:tabres_closed_all} shows the results of the model trained on Yelp population data we collected from Chicago. Although the model is simple, it fits the data well (AUC=$0.674$) and all the variables are statistically significant ($p-value<0.01$).   

We can gain further insight by examining the marginal contributions of each of the factors to the probability of survival. Therefore, we make the following observations from the results:  
\begin{itemize}
\item When the average Yelp rating is higher, the risk of failure gets lower. Higher rated business tend to be more successful. 
\item When the number of reviews is higher, the risk of failure increases.  A business with high number of reviews has been around for a longer time and its risk increases with time.
\item When the average Yelp rating weighted by the number of reviews $rate \times nreview$ increases, the risk of failure gets lower. This is consistent with the theory of Bayesian learning presented in 
\cite{Luca2011}. This is because the average Yelp rating weighted by the number of reviews gives more precise information compared to rating or number of reviews only.
\item The business location makes a difference, some zip codes are riskier than others. This is consistent with previous work on restaurant failure by zip code.
\item The price risk computed for the business price category is also significant similar to the zip code risk.
\end{itemize}

\paragraph{Groupon Adoption Model}
\vspace{1mm}
\noindent
Similar to the survival model, we model the Groupon decision variable {\em isGroupon} as a probit function of a number of business variables from Yelp (as in Table~\ref{tab:vardesc}):
\begin{equation}
isGroupon  = f(rate \times nreview, rate , nreview, gzrisk, gprisk)
\label{eqn:groupon_probit}
\end{equation}
We selected the specification of Equation \ref{eqn:groupon_probit} based on the AIC .
Table~\ref{tab:tabres_groupon_all} shows the results of the model trained on  Yelp population data from Chicago. The model has a good accuracy of AUC = $0.894$, that shows the ability of the model to predict whether a business will decide to make a daily deal based on some business parameters collected from Yelp. Also, the regressors are statistically significant ($p-value<0.01$). 
We can gain further insight by examining the marginal contributions of each of the factors to the probability of offering a daily deal. Therefore, we make the following observations  from the results:  
\begin{itemize}
\item When the average Yelp rating is higher, the probability of daily deal increases. We conjecture that this is due in part to how daily deal sales force selects the business to approach. 
\item When the number of reviews is higher, the probability of daily deal increases.  We conjecture that this is also due in part to how daily deal sales force selects the business to approach.
\item When the average Yelp rating weighted by the number of reviews $rate \times nreview$ increases, the probability of daily deal  gets lower.  This is an indication of a more successful business that would not need to make a deal at Groupon, especially that Groupon takes $50\%$ of the deal revenue \cite{arabshahi}. This results is also consistent with the theory of Bayesian learning presented in \cite{Luca2011}.
\item The business location in terms of zip code makes a difference, some zip codes are more likely to offer daily deals. The higher the number of businesses that make a deal in the same zip code, the higher the chance of a business to adopt a deal. This is part due to completive pressure and in part due to  how daily deal sales team targets geographical areas.  
\item The business price category is also significant as we mentioned before. 
\end{itemize}

\subsection{Joint Model}
We jointly modeled the Groupon and failure models using Equations \ref{eqn:groupon_probit} and \ref{eqn:survival_probit} using a bivariate probit  model. Table~\ref{tab:tabres_aic} shows that the  data is best modeled by a bivariate probit since the log likelihood of the bivariate model differs from that of the two separate equations by more than "1.0". In addition, Tables  \ref{tab:tabres_biprobit_restbar} and \ref{tab:tabres_biprobit_spas} show that for the two largest daily deal categories, the correlation between the unobserved factors "$\rho$" is positive and significant ($0.281$ in Restaurants/Bars, $0.24$ in Beauty/Spas).  As a robustness test, we used the SURE model to compute the correlation between the unobserved errors and tested its significance using the Breusch-Pagan test (correlation=$0.054$ in Restaurants/Bars, $0.060$ in Beauty/Spas). It should be noted that the bivariate probit model operates in the space of latent variables that can be in range [$-\infty,0$] when the dependent variable is "0" and in the range of [$0 , \infty$] when the dependent variable is "1". The residuals are therefore computed in that space and can assume large values. On the other hand, the SURE model operates directly in the space of dependent variable that assume range between "0" and "1". Therefore the residuals are smaller in case of SURE. This explains why the correlation coefficient in case of bivariate probit differs from that in case of SURE model. 

Previous work has shown that restaurants with their high marginal cost, low fixed cost and inability to schedule the arrival of daily deal customers \cite{Gupta} are not well suited to daily deals. On the other hand, Spas with their low marginal cost, high fixed cost and ability to schedule daily deal customers are better suited for daily deals. Conversely, a daily deal is a more  desperate measure for a restaurant than a Spa. This is validated by our results, if a restaurant offers a daily deal with out having any strong  reason to do so, its probability of failing increases more than a Spa that had some motivation for offering a daily deal. 
Similar to what we did in the univariate case, we analyzed the contributions of factors, and we have the following observations:
\begin{itemize}
\item When the ratings weighted by the number of reviews $rate \times nreview$ increases, the risk of failure gets lower. Also, the probability of offering a daily deal gets lower. 
\item Unlike the univariate case, the ratings $rate$ is not a significant factor. However, the weighted ratings are significant. This is consistent with the theory of Bayesian learning presented in  \cite{Luca2011}.
\item As in the univariate case, when the number of reviews is higher, the risk of failure increases.
\end{itemize}

We have also tried other business categories, for example, Health \& Medical, Active Life,..{\em etc}. However, there was a lack of significance of correlation between the unobserved factors for these categories. We conjecture the lack of significance is because we did not  have enough samples in  those categories. In future work, we aim to extend our study to daily deal data from Yipit \cite{yipit} as well as collecting data from  other sources such as city department of revenue to gain a better access to businesses that are not represented well by our current data.

\begin{table}
\begin{tabular}{l l r @{} c c c }
\hline\hline
\multicolumn{6}{c}
{Restaurants and Bars}\\\hline
\multicolumn{2}{l}
{\textbf{Model Name}}
& \multicolumn{2}{l}{\textbf{No. Params}} &  & \textbf{AIC} \\ \hline
isGroupon+isClosed  & & {10} & & &{1976.46}\\
Bivariate Probit & & {11} & & &\textbf{1968.33}\\
\hline
\multicolumn{6}{c}
{Beauty and Spas}\\\hline
\multicolumn{2}{l}
{\textbf{Model Name}}
& \multicolumn{2}{l}{\textbf{No. Params}} & & \textbf{AIC} \\ \hline
isGroupon+isClosed & & {10} & & &{1029.97}\\
Bivariate Probit &  & {11} & & &\textbf{1028.89}\\
\hline\hline
\end{tabular}
\vspace{-2mm}
\caption{AIC for univariate versus bivarite probit}
\vspace{-4mm}
\label{tab:tabres_aic}
\end{table}

\def\onepc{$^{\ast\ast}$} \def\fivepc{$^{\ast}$}
\def\tenpc{$^{\dag}$}
\def\legend{\multicolumn{4}{l}{\footnotesize{Significance levels
:\hspace{1em} $\dag$ : 10\% \hspace{1em}
$\ast$ : 5\% \hspace{1em} $\ast\ast$ : 1\% \normalsize}}}
\vspace{-1mm}

\begin{table*}[ht]
\parbox[c]{.5\textwidth}{
\centering
\caption{Restaurants and Bars: Bivariate Probit}
\vspace{2mm}
\label{tab:tabres_biprobit_restbar}
\begin{tabular}{l r @{} l c }\hline\hline
\multicolumn{4}{c}{Bivariate Probit}\\
\hline
\multicolumn{1}{c}
{\textbf{Variable}}
& \multicolumn{2}{c}{\textbf{Coefficient}}  & \textbf{(Std. Err.)} \\ \hline
\hline \multicolumn{4}{c}{Equation 1 : isGroupon} \\ \hline
Groupon Pricerisk  &  13.978&\onepc  & (2.122)\\
Groupon Ziprisk  &  5.715&\onepc  & (0.535)\\
Rating $\times$ Reviews  &  -0.002&\onepc  & (0.001)\\
Reviews  &  0.010&\onepc  & (0.002)\\
Rating  &  0.180&\fivepc  & (0.073)\\
Intercept  &  -3.385&\onepc  & (0.293)\\
\hline \multicolumn{4}{c}{Equation 2 : isClosed} \\ \hline
Fail Ziprisk  &  8.791&\onepc  & (1.021)\\
Fail Pricerisk  &  12.928&\onepc  & (3.541)\\
Rating $\times$ Reviews  &  -0.002&\fivepc  & (0.001)\\
Reviews  &  0.007&\tenpc  & (0.004)\\
Rating  &  -0.013&  & (0.070)\\
Intercept  &  -4.202&\onepc  & (0.495)\\
\hline \multicolumn{4}{c}{Equation 3 : Joint} \\ \hline
athrho  & \textbf{0.288}&\onepc  & (0.09161)\\  
rho  & \textbf{0.281}&  & (.08440)\\  
\hline
\multicolumn{4}{c}{SURE: Breusch-Pagan test of independence}\\
\hline
Correlation  & \multicolumn{3}{c}{\textbf{0.0543}\onepc}\\
\hline
\legend
\end{tabular}
}
\parbox[c]{.5\textwidth}{
\centering
\caption{Beauty and Spas: Bivariate Probit}
\vspace{2mm}
\label{tab:tabres_biprobit_spas}
\begin{tabular}{l r @{} l c }\hline\hline
\multicolumn{4}{c}{Bivariate Probit}\\
\hline
\multicolumn{1}{c}
{\textbf{Variable}}
& \multicolumn{2}{c}{\textbf{Coefficient}}  & \textbf{(Std. Err.)} \\ \hline
\hline \multicolumn{4}{c}{Equation 1 : isGroupon} \\ \hline
Groupon Pricerisk  &  17.203&\onepc  & (3.326)\\
Groupon Ziprisk   &  4.460&\onepc  & (0.373)\\
Rating $\times$ Reviews  &  -0.004&\onepc  & (0.001)\\
Reviews  &  0.024&\onepc  & (0.006)\\
Rating  &  -0.087&  & (0.059)\\
Intercept  &  -2.613&\onepc  & (0.288)\\
\hline \multicolumn{4}{c}{Equation 2 : isClosed} \\ \hline
Fail Ziprisk  &  5.199&\onepc  & (1.556)\\
Fail Pricerisk  &  8.906&\fivepc  & (4.538)\\
Rating $\times$ Reviews  &  -0.007&\fivepc  & (0.003)\\
Reviews  &  0.026&\onepc  & (0.010)\\
Rating  &  0.103&  & (0.090)\\
Intercept  &  -4.201&\onepc  & (0.719)\\
\hline \multicolumn{4}{c}{Equation 3 : Joint} \\ \hline    
athrho  &  \textbf{0.25}&\tenpc  & (0.1433)\\
rho  &  \textbf{0.246}&  & (0.1346)\\    
\hline
\multicolumn{4}{c}{SURE: Breusch-Pagan test of independence}\\
\hline
Correlation  & \multicolumn{3}{c}{\textbf{0.060}\onepc}\\
\hline
\legend
\end{tabular}
}
\end{table*}

\section{Conclusion and Future Work}
\label{future}

In this work, we studied  whether daily deal adoption signals additional information about the business. For the two largest daily deal categories restaurants and spas, we found that the unobserved factors that contribute to a business decision to offer a daily are positively correlated with unobserved factors that contribute to a business failure. Restaurants had a higher correlation while spas had a lower correlation. These results indicate that daily deal provide a strong signal of business survival for restaurants and to a lesser extent for spas. 
Our results also show that social media sites such as Yelp provide a rich set of information that can be used to model business. In particular, we found that consistent with Bayesian learning theory, the rating of business weighted by the number of reviews provides a statistically significant predictor of business failure.{\em Ceteris paribus}, a business with a high number of positive reviews has higher odds of survival.

In future, we plan to extend our work in multiple directions. The first direction is to consider other business categories. In this work, we were limited  to the daily deal data from  \cite{byers_a}.   Second, we plan to consider daily deal providers other than Groupon. In the case of Groupon, we modeled the marketers decision to use daily deal  as a binary decision. We also modeled the business failure as a binary decision. This allowed us to leverage the framework of bivariate probit to jointly model daily deal and business failure. In the more general setting, the marketer can choose to offer a daily deal through a number of providers or not to offer a deal. In the general case, the marketer's choice is best modeled as a multinomial. We plan to investigate techniques from Multilevel Multiprocess Models (MLMP)  \cite{LillardBrianWaite}. 
The data needed to undertake these two directions can be obtained from daily deal aggregators such as Yipit \cite{yipit}

We also plan to address the dynamics of Groupon adoption and business failure. In particular, we plan to test whether daily deal  adoption is stationary, whether business failures are stationary and whether the two time series are co-integrated \cite{Hamilton94}. This will allow us to investigate whether there is a long term equilibrium between daily deal adoption and business failure. 

Last but not least, we plan to investigate the causal impact of Groupon on metrics other than business failure. We plan to investigate the Groupon impact on sales, number of orders and order size. This will help us better understand how the daily deals impact revenue: a) do they impact revenue through a change in the number of orders. or b) do daily deals have a stronger impact on order size or c)  do daily deals have an impact on both number and size of orders.
We will also look at using instruments \cite{Green6}  to get an estimate of the casual impact of Groupon. We are investigating Groupon sales force as a potential instrument. 

\bibliographystyle{abbrv}
\bibliography{aof}
\end{document}